\documentclass[useAMS,usenatbib]{mnras}
\usepackage{graphicx}
\usepackage{amsmath}
\usepackage{amssymb}
\usepackage{subfigure}
\usepackage{url}
\usepackage{multirow}
\usepackage{grffile}

\newcommand{\km}{\rm\thinspace km}

\newcommand{\Hz}{\rm\thinspace Hz}

\newcommand{\Msun}{\hbox{$\rm\thinspace M_{\odot}$}}

\newcommand{\keV}{\rm\thinspace keV}

\newcommand{\rg}{\rm\thinspace $r_\mathrm{g}$}

\hypersetup{draft}

\title[Illumination of neutron star accretion discs]{On the illumination of neutron star accretion discs}
\author[D. R. Wilkins]{D. R. Wilkins\thanks{E-mail: dan.wilkins@stanford.edu}\thanks{Einstein Fellow}\\
Kavli Institute for Particle Astrophysics and Cosmology, Stanford University, 452 Lomita Mall, Stanford, CA 94305, USA \\
}
\begin{document}

\date{Accepted 2017 December 5. Received 2017 December 5; in original form 2017 May 1}

\pagerange{\pageref{firstpage}--\pageref{lastpage}} \pubyear{2017}

\maketitle

\label{firstpage}

\begin{abstract}
The illumination of the accretion disc in a neutron star X-ray binary by X-rays emitted from (or close to) the neutron star surface is explored through general relativistic ray tracing simulations. The applicability of the canonical suite of relativistically broadened emission line models (developed for black holes) to discs around neutron stars is evaluated. These models were found to describe well emission lines from neutron star accretion discs unless the neutron star radius is larger than the innermost stable orbit of the accretion disc at 6\rg\ or the disc is viewed at high inclination, above 60\,deg where shadowing of the back side of the disc becomes important. Theoretical emissivity profiles were computed for accretion discs illuminated by hotspots on the neutron star surfaces, bands of emission and emission by the entirety of the hot, spherical star surface and in all cases, the emissivity profile of the accretion disc was found to be well represented by a single power law falling off slightly steeper than $r^{-3}$. Steepening of the emissivity index was found where the emission is close to the disc plane and the disc can appear truncated when illuminated by a hotspot at high latitude. The emissivity profile of the accretion disc in Serpens X-1 was measured and found to be consistent with a single unbroken power law with index $q=3.5_{-0.4}^{+0.3}$, suggestive of illumination by the boundary layer between the disc and neutron star surface. 

\end{abstract}


\section{Introduction}
Neutron stars represent unique laboratories in which the equation of state of ultradense matter can be studied. The mass-radius relation of the neutron star is a sensitive probe of this. Neutron star radius measurements, in particular through X-ray observations, can discriminate between different models of the ultradense material in the stellar interior, observing the thermal spectrum of X-ray bursters \citep{guver+2010} and quiescent X-ray transients \citep{guillot+2013}.

Where neutron stars are found in X-ray binaries and accrete from their stellar companion, measurements of the disc of material accreting onto the neutron star may also provide constraints on its radius and hence the equation of state of the ultradense matter in the neutron star. A characteristic reflection spectrum is produced where the accretion disc is illuminated by an external X-ray source \citep{ross_fabian,garcia+2010,garcia+2011,garcia+2013}, for example emission generated in a hotspot on the surface of the neutron star or in the boundary layer in which material orbiting at the Keplerian velocity on the inner edge of the accretion disc is decelerated to accrete onto the more slowly rotating neutron star, or emission from a magnetised corona associated with the disc.

Emission lines are produced from the disc including the iron K$\alpha$ fluorescence line at 6.4\keV. While emission lines may be narrow in the rest frame of the emitting material, the Doppler shift produced by the orbital motion of the disc material as well as gravitational redshifts in the strong gravitational potential in close proximity to the neutron star surface result in lines being broadened, exhibiting a characteristic redshifted wing \citep{fabian+89}. The extent of the redshifted wing of the line reveals how deep into the gravitational potential the accretion disc extends and hence if the inner edge of the accretion disc can be determined, the neutron star radius can be constrained to lying within this radius. Although they are less prominent than in AGN and black hole X-ray binaries due to additioal continuum emission from the boundary layer, relativistically broadened iron K$\alpha$ emission lines have been detected from the accretion discs around a growing number of neutron stars in \textit{XMM-Newton}, \textit{Suzaku} and, more recently, \textit{NuSTAR} \citep{bhatt+2007,dai+2009,disalvo+2009,iaria+2009,papitto+2009,egron+2013,miller+2013}.

High X-ray count rates from neutron star X-ray binaries in the past has complicated the accurate measurement emission line profiles due to photon pile-up in CCD detectors. \citet{ng+2010} suggest that iron K$\alpha$ lines from neutron stars may in fact be intrinsically narrow and symmetric, only appearing to be broadened due to the distortions to the spectrum induced by photon pile-up. \citet{miller+2010}, on the other hand, present simulations that suggest that photon pile-up falsely \textit{narrows} rather than broadens lines. It is therefore greatly advantageous to study the X-ray emission from neutron stars using detectors that do not suffer photon-pileup. Gas spectrometers point to relativistically broadened lines from neutron star accretion discs \citep{lin+2010,cackett+2012} but are limited by their low spectral resolution. \textit{NuSTAR} \citep{nustar} however has been revolutionary for the study of neutron stars with its CZT detectors providing sufficient spectral resolution to measure the profile of the lines while being unaffected by photon pile-up. Observations of neutron star X-ray binaries by \textit{NuSTAR} have yielded definitive detections of relativistically broadened lines from neutron star accretion discs \citep{miller+2013}.

In order to measure the inner edge of the accretion disc from the profile of relativistically broadened emission lines, it is important to understand the pattern of illumination of the accretion disc by the primary X-ray source, characterised by its \textit{emissivity profile}. The emissivity profile can be degenerate with the inner radius of the accretion disc when models are fitted to broad emission lines, for example weak illumination of the inner disc can be incorrectly interpreted as truncation of the disc \citep{emis_spin_degen}. A thorough underdstanding of the illumination of neutron star accretion discs and how reflection from the disc can constrain the neutron star equation of state is particularly opportune with the advent of the \textit{Neutron Star Interior Composition Explorer} or \textit{NICER} \citep{nicer} which will measure precisely the emission line profiles in neutron star X-ray binaries with a combination of a large collecting area and detectors that will not suffer photon pile-up.

In the case of relativistically broadened lines in AGN and black hole binaries, it has proven possible to directly measure the emissivity profile of the accretion disc by fitting the contribution to the observed line profile from each radius on the disc, distinguished by varying Doppler shifts and gravitational redshift \citep{1h0707_emis_paper, cygx1_spin}. As well as enabling the measurement of the inner radius of the disc, comparing the measured accretion disc emissivity profile to theoretical predictions derived from general relativistic ray tracing simulations allows the geometry and location of the illuminating X-ray source to be inferred \citep{understanding_emis_paper}.

In this work, the illumination of neutron star accretion discs are studied with the aim of understanding the correct profile that should be assumed when measuring the inner disc radius and understanding the nature of the primary X-ray source that illuminates the disc in accreting neutron star X-ray binaries. It is first explored whether the canonical relativistic broad line prescriptions are appropriate for modelling neutron star spectra given the potential shadowing of part of the disc by the star surface, before theoretical emissivity profiles are computed for neutron star accretion discs illuminated by hot spots as well as the entire star surface or boundary layer. Line profiles emerging from non-axisymmetric illumination of the disc due to shadowing from the star are studied and the results are applied to \textit{NuSTAR} observations of the neutron star low-mass X-ray binary Serpens X-1 (Ser~X-1).

\section{Verification of the Line Description}
\label{line_desc.sec}
The standard relativistically-broadened emission line models assume that the illumination of the accretion disc by the primary X-ray source is axisymmetric and no part of it is obscured, and hence that the illumination pattern can be described by a radial emissivity profile. Before computing the emissivity profiles expected for the accretion discs around neutron stars, the validity of this description is verified given that the solid star surface can potentially obscure the line of sight to the rear part of the disc.

The spacetime around the neutron star is modelled by the Kerr geometry, albeit in the limit of slow rotation. Since neutron stars are observed to be slowly rotating ($a \lesssim 0.1\,GM/c^2$, which corresponds to a spin frequency less than 250\,Hz), we will set the spin parameter $a=0$ whereupon the Kerr geometry reduces to the Schwarzschild geometry for the spacetime outside a non-rotating point source and the accretion disc extends in as far as the innermost stable circular orbit at 6\rg.

An image plane was constructed 10,000\rg\ from the centre of the star, as in \citet{lag_spectra_paper}. The rays passing perpendicular through this plane represent those that will be observed by a telescope. A regular grid of perpendicular rays were traced backwards towards the neutron star until they reach either the star surface or intersect the accretion disc, assumed to lie in the equatorial plane. Recording the location the ray lands on the disc as well and calculating its redshift between the plane and disc allows the image of the accretion disc, subject to relativistic effects around the neutron star, to be constructed.

\begin{figure*}
\centering
\subfigure[$r_\star = 4.5$\rg] {
\includegraphics[width=80mm,angle=180]{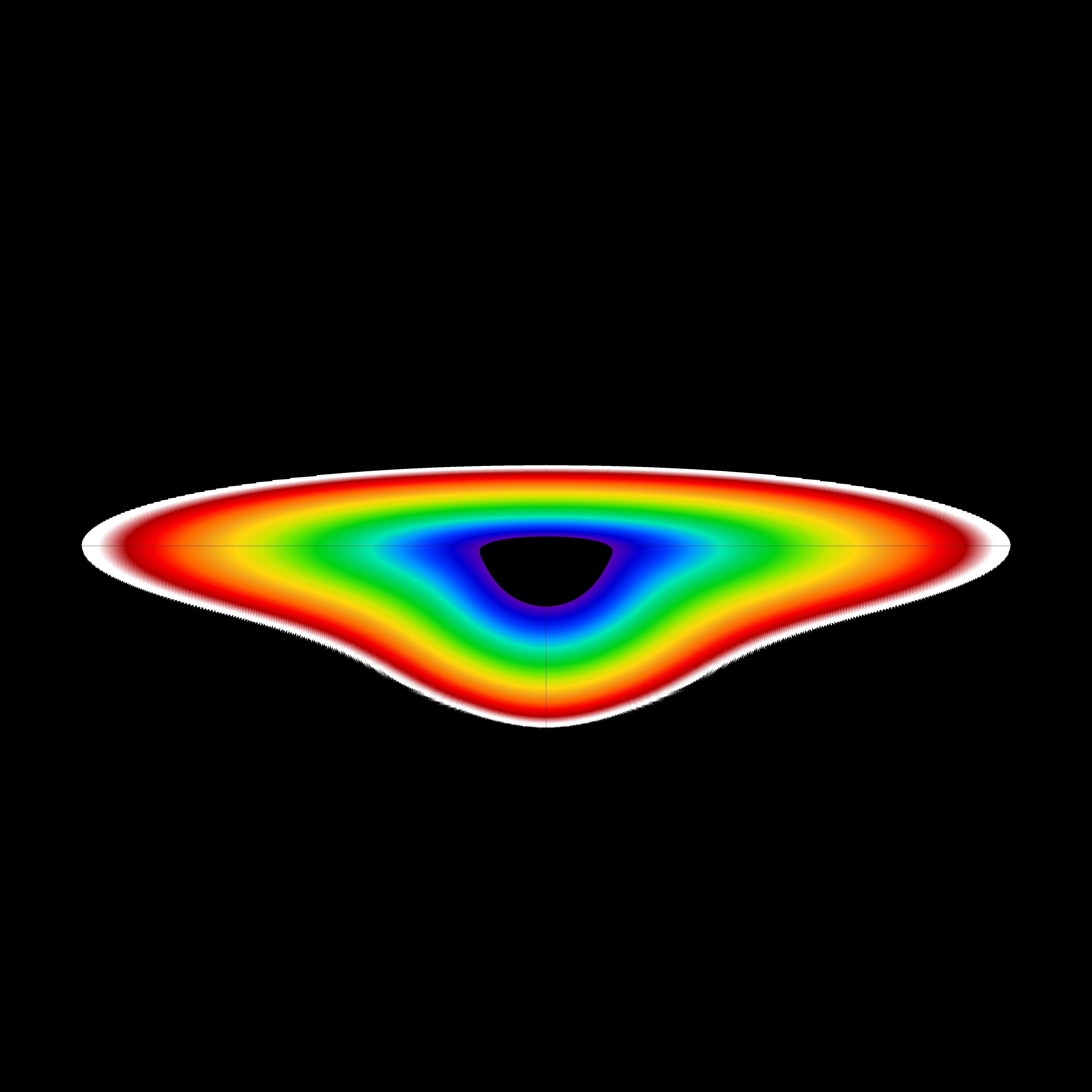}
\label{discplane.fig:4_5}
}
\subfigure[$r_\star = 10$\rg] {
\includegraphics[width=80mm,angle=180]{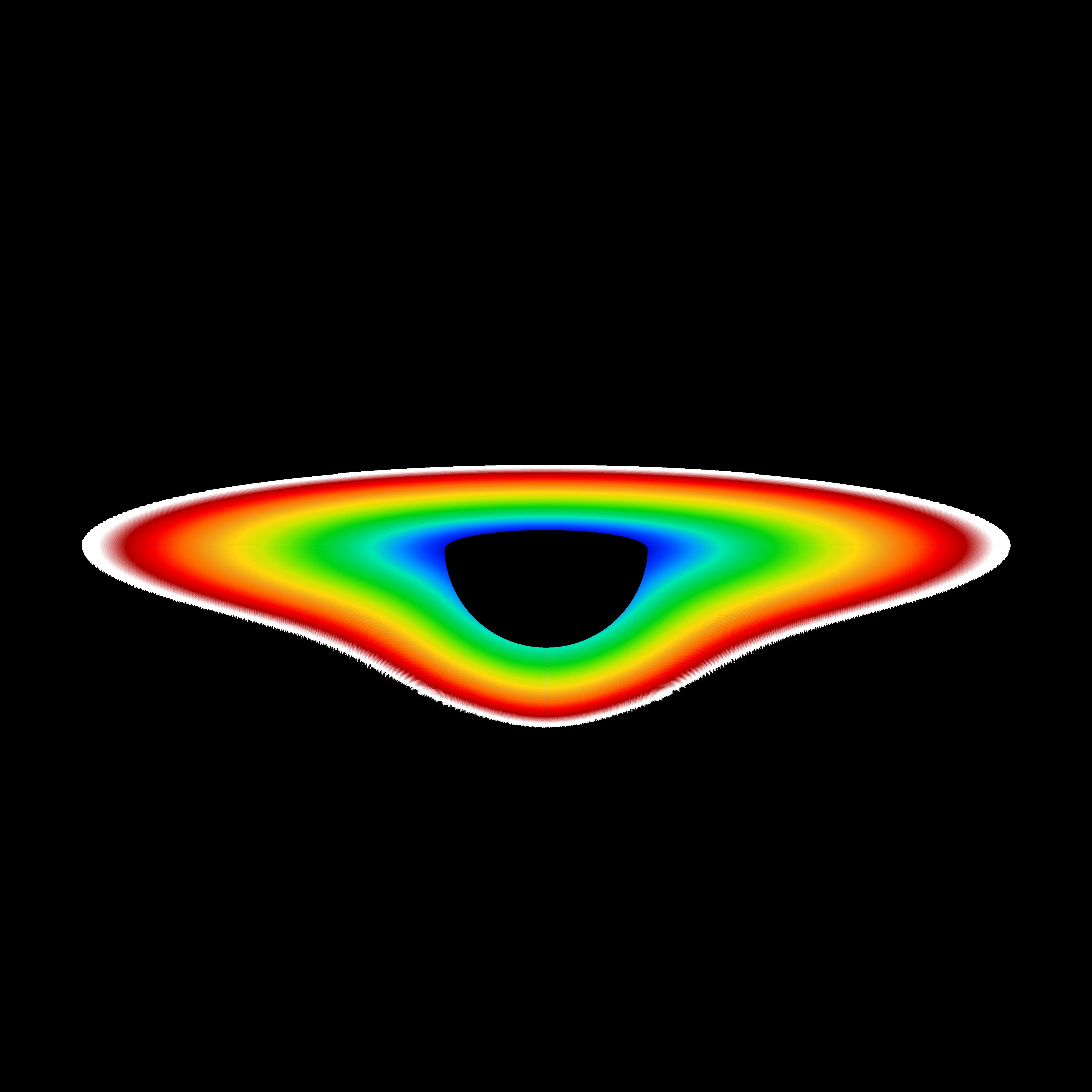}
\label{discplane.fig:10}
}
\caption{Ray traced images of the accretion discs around neutron stars of radius \subref{discplane.fig:4_5} 4.5\rg\ and \subref{discplane.fig:10} 10\rg\ (exaggerated case) to illustrate the obscuration of the back side of the accretion disc by the solid star surface. The line of sight is inclined at an extremal 80\,deg to the normal to the disc. Colour shading corresponds to the radial co-ordinate on the accretion disc from which the rays originate.}
\label{discplane.fig}
\end{figure*}

For illustration, Fig.~\ref{discplane.fig} shows ray traced images of the accretion discs around a neutron star of radius 4.5\rg\ (around 10\km\ assuming mass 1.4\Msun) as well as an exaggerated star radius 10\rg. Shading represents the radius of emission measured in the plane of the accretion disc to show which radii are masked and also to illustrate the bending of light from the back side of the disc around the top of the neutron star. Each patch of the disc is assumed to emit isotropically in its own rest frame. Each patch emits line flux proportional to its proper area, weighted by a power law emissivity profile according to its radius, following $r^{-3}$. 

\begin{figure}
\centering
\includegraphics[width=80mm]{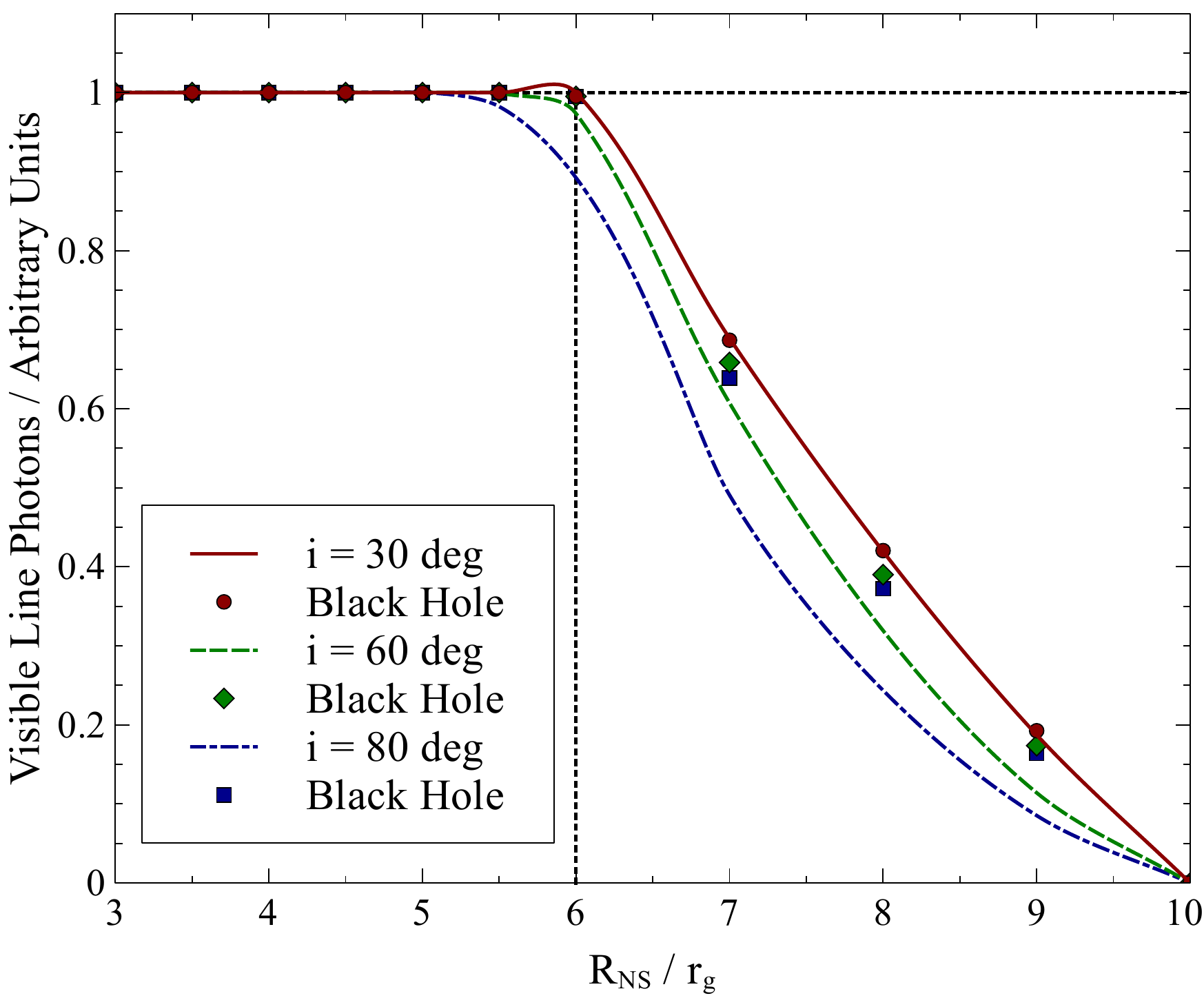}
\caption{The relative emission line count rate received from within a radius of 10\rg\ on the accretion disc as a function of the radius of the neutron star, quantifying the shadowing of the accretion disc by the star. The lines show the count rate from the neutron star accretion discs and the corresponding coloured points show the count rate from a black hole accretion disc truncated at the same radius (though all discs are truncated at 6\rg, the innermost stable orbit in the Schwarzschild geometry).}
\label{discarea.fig}
\end{figure}

In order to measure the shadowing of the back side of the disc by the neutron star, Fig.~\ref{discarea.fig} shows the fraction of line photons emitted within a radius of 10\rg\ (where the redshifted wing of the line originates) that can be seen by the observer. Clearly, a greater fraction of the disc would be masked by the neutron when observed at higher inclination (closer to edge-on). For inclinations 60\,deg and lower, even when the neutron star is touching the inner edge of the disc at 6\rg, only 3 per cent of line photons from within 10\rg\ are blocked. When the inclination is as high as 80\,deg, 10 per cent of the photons are blocked when the star radius is 6\rg. For radii below 6\rg, even with an inclination as high as 80\,deg, there is not significant masking of the inner disc by the neutron star surface, aided by the fact that rays from the back side of the disc that would classically be blocked can be bent around the star in the strong gravitational field.

In each case, the fraction of line photons received from the neutron star accretion disc is compared to that received from an accretion disc around a black hole, truncated at the same radius (shown as points in Fig.~\ref{discarea.fig}). Shadowing by the neutron star surface does, however, starts to have a more significant effect once the neutron star surface is touching the inner edge of the disc and the disc is truncated. Once the neutron star star radius is greater than 6\rg, a discrepancy is evident between the shadowed neutron star accretion disc and what might be expected simply by truncating the black hole accretion disc at the radius corresponding to the stellar surface. For an inclination of 60\,deg, the masking of the back side of the disc by the neutron star leads to a further 10 per cent reduction in disc flux compared to the truncated black hole disc, increasing to a 35 per cent reduction for an inclination of 80\,deg.

\begin{figure*}
\centering
\includegraphics[width=175mm]{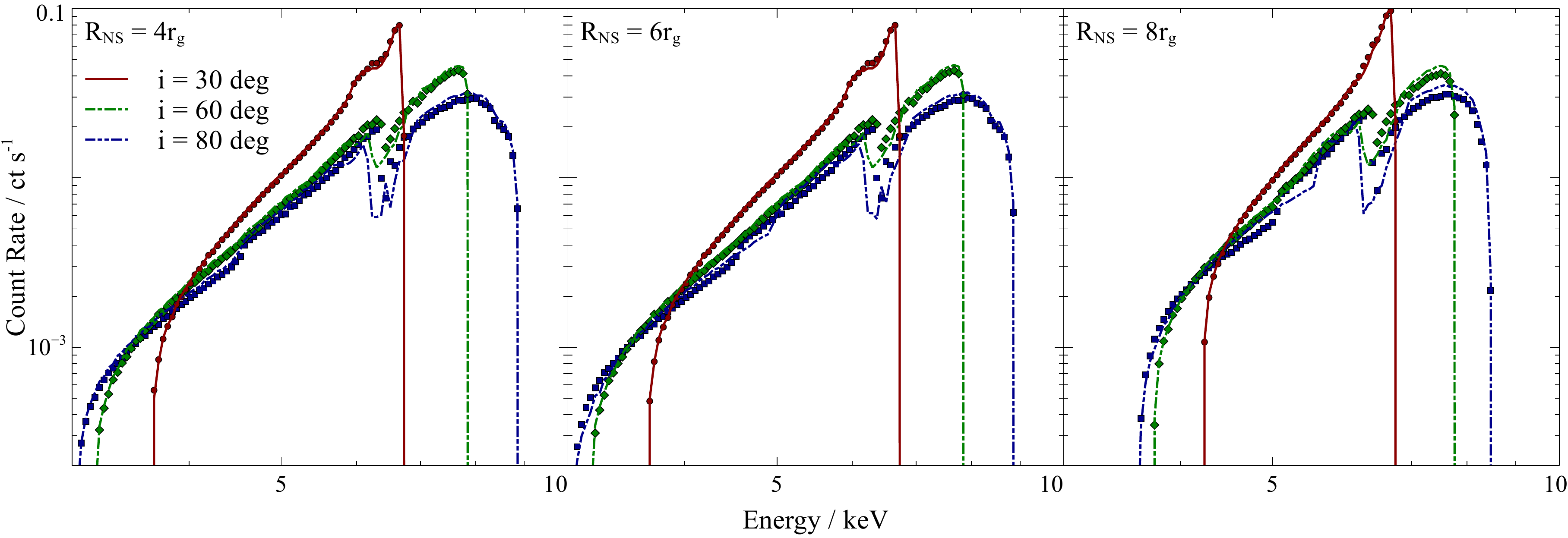}
\caption{Relativistically broadened emission line profiles expected from accretion discs around neutron stars of different radius. Lines are shown for lines of sight at inclination 30, 60 and 80\,deg from the normal to the disc. Lines show the line profiles from neutron star accretion discs while the corresponding coloured points show the line profile from a black hole accretion disc truncated at the same radius (though not extending within 6\,rg, the innermost stable orbit) but with no shadowing from the neutron star surface. At high inclination, when the neutron star radius is 6\rg\ or greater, the neutron star blocks some of the line photons that would be observed from the back side of the disc. These blocked photons appear around 6\keV\ for a 6.4\keV\ line. In each case, the emissivity of the disc falls off as $r^{-3}$.}
\label{lines.fig}
\end{figure*}

Fig.~\ref{lines.fig} shows the broad emission line profiles that would be expected from these discs. The energy at which line photons from each patch are observed is determined from the rays between each patch and the image plane with this redshift factor also determining the specific intensity of each part of the disc that is measured by the observer relative to that measured in the frame of the emitter, following the standard procedure.

These line profiles show the effect of masking the disc on observations. The neutron star masks the back side of the accretion disc where material is travelling approximately transversely to the line of sight, so emission is subject only to gravitational redshift and transverse Doppler shift. The photons lost from the 6.4\keV\ iron K$\alpha$ fluorescence line appear around 6\keV\ and the subtle reduction between the neutron star disc lines (plotted as lines) and the equivalent lines from a truncated black hole disc (points) can be seen. 

These figures show that only in the case of a large neutron star radius, greater than the 6\rg\ innermost stable orbit, or when the accretion disc is observed at an inclination greater than 60\,deg does the obscuration of the back side of the accretion disc by the star start to have a significant impact the profile of the broad emission line. In this case, there is a deficit in the mildly redshifted emission originating from behind the star. Comparing the line profiles to those produced from black hole accretion disc models, however, shows that even for the largest neutron stars whose discs are observed at high inclination, the most blue- and redshifted emission, arising from the sides of the accretion disc that are approaching or receding from the observer are not impacted. This means that the parameters that are often derived from the broad emission line profiles such as the inner radius of the disc, used to measure the spin as well as the emissivity profile of the accretion disc (which are constrained by the extent and shape of the redshifted wing) and the accretion disc inclination (inferred from the most blueshifted emission in the line) can be accurately measured applying the canonical suite of black hole emission line models.

\section{Accretion Disc Emissivity Profiles}
The \textit{emissivity profile} of the accretion disc is defined as the reflected flux emitted per unit proper area of the accretion disc as measured in the rest frame of the emitting material \citep{1h0707_emis_paper,understanding_emis_paper,laor-91}. When fitting model emission lines to observed X-ray spectra, it is typically parametrised as a power law, $\epsilon(r)\propto r^{-q}$, although once- or twice-broken power laws are often required to fit the observed spectra of black hole binaries and AGN. By definition, the emissivity profile is a function of radius only (it is assumed that the disc is illuminated axisymmetrically) and hence represents the angle-averaged emission from each radius on the disc.

Emissivity profiles of accretion discs around neutron stars are calculated following the method of \citet{understanding_emis_paper}. Rays are traced from the X-ray source by numerical integration of the null geodesic equations in the spacetime around the neutron star, until they reach the equatorial plane in which the accretion disc is assumed to lie. The emissivity profile is calculated by counting the total energy that is incident in radial bins on the disc (the product of the number of rays and the blue- or redshift of each ray at the disc relative to its emission point). The energy in each bin is divided by the area of the bin, measured in the frame of the orbiting material in the disc, allowing for relativistic corrections due to the orbital motion of the disc material. Moreover, relativistic time dilation between the X-ray source and observers at locations on the accretion disc in the stronger gravitational field, closer to the neutron star is important since each ray may be taken to represent emission of photons at a constant rate as far as the emitter is concerned. This results in the relative enhancement of the photon arrival rate seen by observers closer to the centre of the potential for whom proper time elapses more slowly.

\subsection{Point Sources, Hotspots \& Belts of Emission}
\begin{figure}
\centering
\includegraphics[width=85mm]{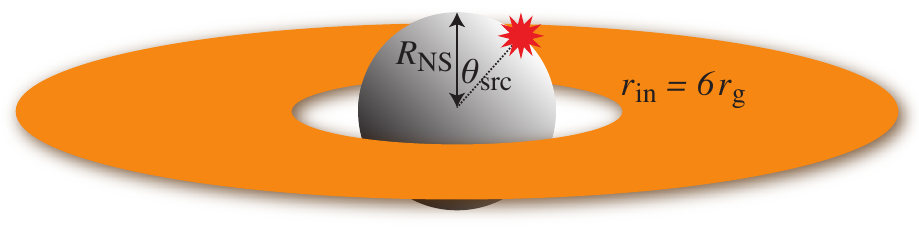}
\caption{The geometry assumed in simulations of neutron star accretion discs illuminated by a point source (\textit{i.e. a hot spot}) on the neutron star surface. Due to the axisymmetry of the spacetime, this is equivalent to illumination by a band of emission at the same polar angle, $\theta_\mathrm{src}$.}
\label{pointsource.fig}
\end{figure}

The geometry that is simulated of a neutron star accretion disc illuminated by a point source is shown in Fig.~\ref{pointsource.fig} with two parameters; the radius of the neutron star, $R_\mathrm{NS}$ and the latitude of the source on the star surface, characterised by the angle from the pole, $\theta_\mathrm{src}$. In order to compute the illumination of the accretion disc by a point source, rays were started at regular steps in the polar angles, $\cos\alpha$ and $\beta$ measured in the frame of the emitter, to model an isotropic point source emitting equal power into equal solid angle in its own rest frame.

Emission from a localised point source might be seen if a hotspot is produced on the surface of the neutron star, for example if the accreting material is funnelled onto the poles of the star's magnetic dipole moment and the period of rotation of the star is much longer than the duration of an observation

Due to the axisymmetric nature of the Schwarzschild (and Kerr) spacetime, the emissivity profile (which is angle-averaged at each radius) produced by a point source is equivalent to that produced by a ring or band of emission around the surface of the star at the same $\theta_\mathrm{src}$ co-ordinate. While the spacetime might be axisymmetric, the observation of the disc breaks this symmetry with different Doppler shifts in the reflected X-rays that are observed from approaching and a receding sides. If the hot spot were not located on the pole of the neutron star, one side of the disc would be preferentially illuminated, resulting in a deviation of the observed emission line profile from the canonical black hole line models. Accreting neutron stars, however, are typically seen to spin with frequencies of 100 to 700\Hz\ \citep{watts_ns_review}, hence the rotation period is much shorter than the timescale of a typical observation. The time-averaged spectrum over the course of an observation will represent the illumination of the disc by an extended ring of emission as the hotspot is translated around the star (assuming that the rotation axis of the star is aligned with the rotation axis of the inner regions of the accretion disc), hence the axisymmetric formulation of the emissivity profile and the canonical line models remain valid.

An extended belt of X-ray emission could also be produced if the X-ray continuum emission that illuminates the disc in order to produce the observed emission lines arises from a boundary layer where the rapidly orbiting material on the inner edge of the accretion disc, travelling at the local Keplerian velocity, must be decelerated in order to accrete onto the surface of the more slowly rotating neutron star. Through the dissipation of the excess angular momentum possessed by material from the inner orbit of the disc, the boundary layer is expected to be the dominant source of high energy radiation in accretion neutron star systems as the hot gas produces an X-ray continuum through Comptonisation \citep{popham_sunyaev}. Such a boundary layer must exist whenever material must decelerate from the inner edge of the accretion disc (where it will be orbiting at half the speed of light if this is the innermost stable orbit). If the surface of the star is close to the inner edge of the disc, one would expect the disc structure to deviate greatly from the standard Shakura-Sunyaev picture of a geometrically thin, optically thick disc orbiting at the Keplerian velocity at each radius, thus the reflection spectrum and Doppler shifts thereof would not be well described by the standard suite of broad emission line models, though such considerations are beyond the scope of this work which concentrates on the illumination pattern of accretion discs in the standard reflection scenario.

In these simulations, the X-ray source is assumed to be stationary. In reality, the the source will rotate with the surface of the neutron star. In the limit that the surface velocity is much less than the speed of light (which it will be in the limit of the Schwarzschild metric), source motion has no significant effect on the illumination pattern cast over the accretion disc. Only when the velocity is a significant fraction of the speed of light does relativistic beaming come into effect. Even in the limit of rapid rotation close to $0.5c$, \citet{understanding_emis_paper} show that relativistic beaming of source photons causes only subtle flattening of the emissivity profile at disc radii closest to the emitter.

\begin{figure*}
\centering
\subfigure[$\theta_\mathrm{src} = 0$\,deg] {
\includegraphics[width=85mm]{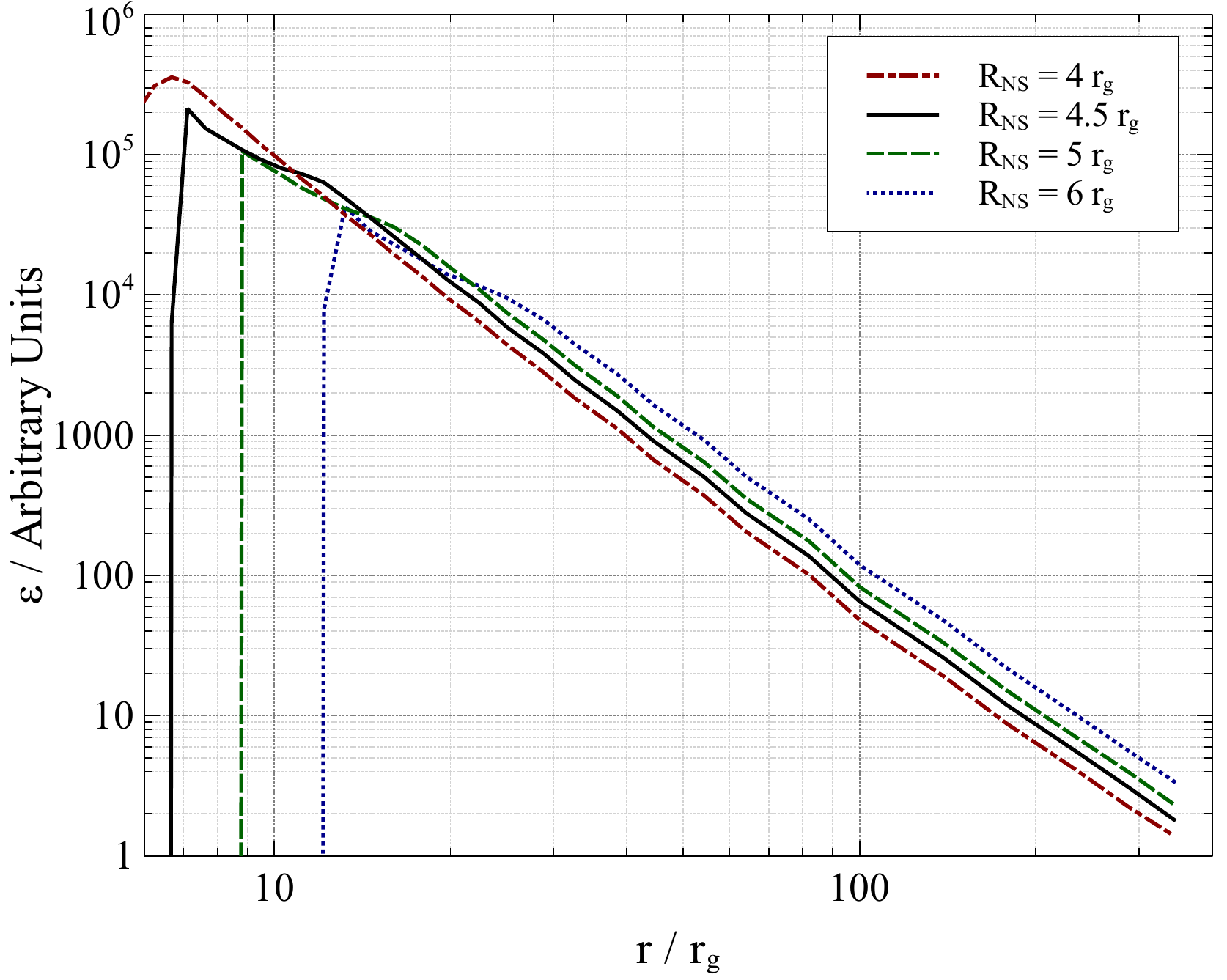}
\label{emis_point.fig:theta0}
}
\subfigure[$\theta_\mathrm{src} = 80$\,deg] {
\includegraphics[width=85mm]{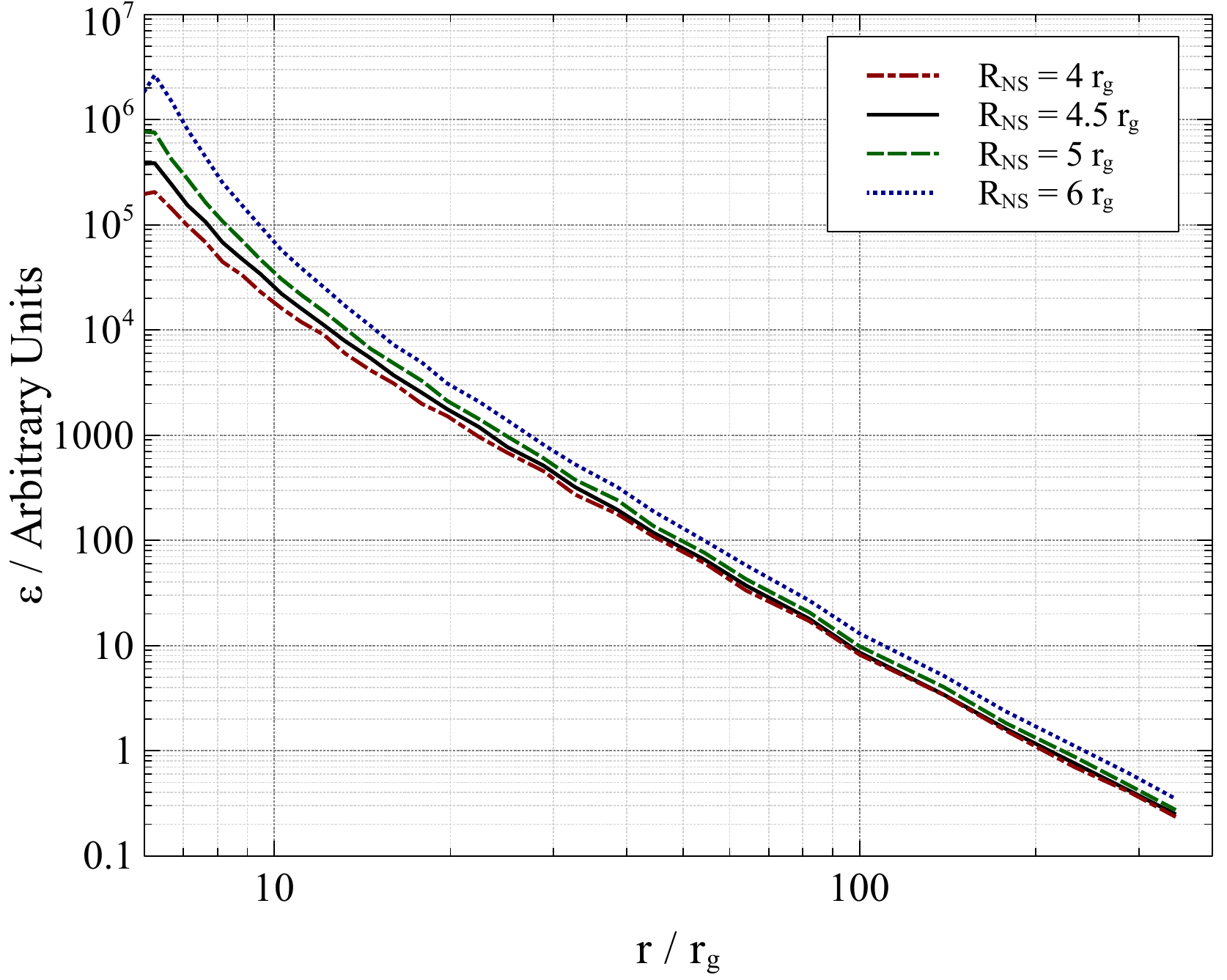}
\label{emis_point.fig:theta80}
}
\caption{Expected emissivity profiles of neutron star accretion discs illuminated by point sources \subref{emis_point.fig:theta0} on the pole of the star and \subref{emis_point.fig:theta80} on the surface of the star closer to the equatorial plane. The neutron star is assumed to be non-spinning, hence the accretion disc is truncated at 6\rg\ from the centre of the star. Point sources on the surfaces of neutron stars with varying radius ($R_\mathrm{NS}$) are considered.}
\label{emis_point.fig}
\end{figure*}

Fig.~\ref{emis_point.fig} shows the calculated emissivity profiles for neutron star accretion discs illuminated by an isotropic point source on the surface of the star, representing an X-ray emitting hot spot and indistinguishable from a belt of emission at the same latitude around the surface. Profiles are shown for a range of radius of the neutron star ($R_\mathrm{NS}$) and for a source located on the pole of the star ($\theta_\mathrm{src} = 0$) and close to the equator ($\theta_\mathrm{src} = 80$\,deg). It can be seen that in both the cases of a hot spot or belt located on the pole of the star and closer to the equatorial plane that the emissivity profile is consistent with constant power law index, approximately $q=3$, over the accretion disc with the exception of X-ray sources close to the equator, where in the case of a larger radius star (at $R_\mathrm{NS} = 6$\rg), the close proximity of the source to the inner edge of the accretion disc (in fact for a 6\rg\ neutron star the source is right on the inner edge of the accretion disc) causes a pronounced steepening of the emissivity profile (reaching a power law index of $q=6$) within 10\rg\ on the disc.

The form of the emissivity profile around a neutron star is equivalent to that around a black hole by such a source, as explored by \citet{understanding_emis_paper}, though without the steepened inner section of the emissivity profile (at $r<5$\rg) since we here consider accretion discs around non-spinning neutron stars where the disc extends in only as far as $r=6$\rg.

\begin{figure*}
\centering
\subfigure[$R_\mathrm{NS} = 4.5$\rg] {
\includegraphics[width=55mm]{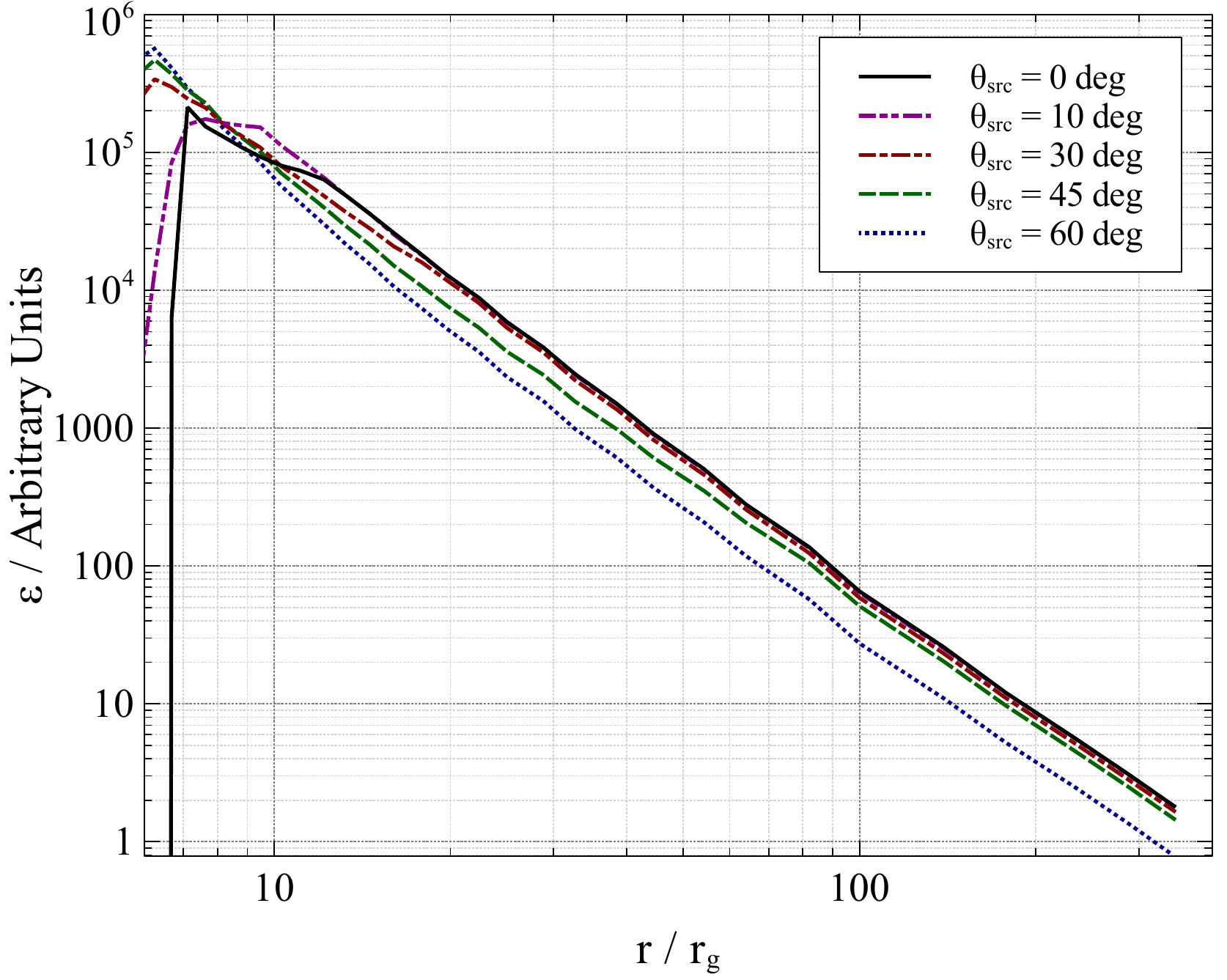}
\label{emis_theta.fig:r45}
}
\subfigure[$R_\mathrm{NS} = 5$\rg] {
\includegraphics[width=55mm]{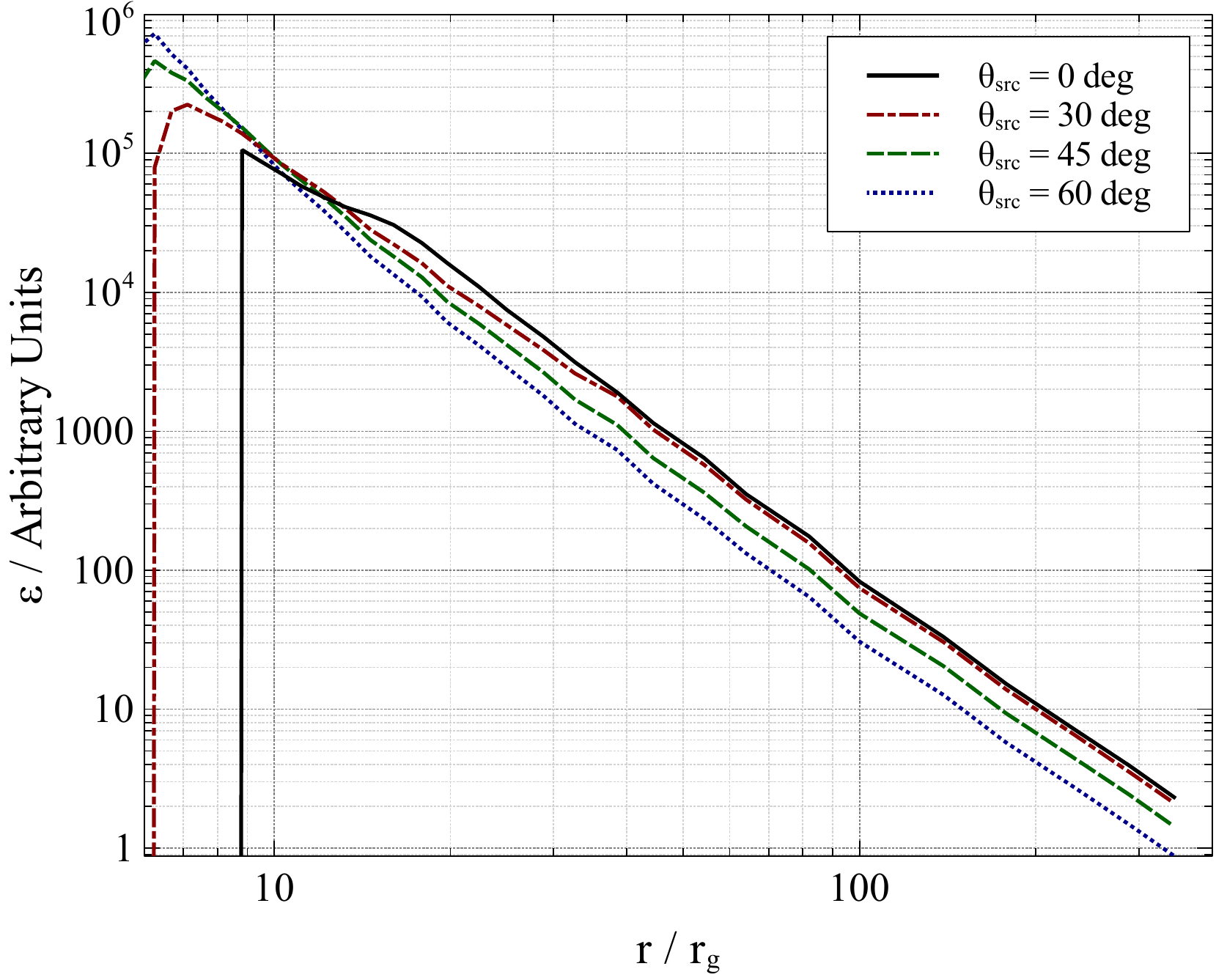}
\label{emis_theta.fig:r5}
}
\subfigure[$R_\mathrm{NS} = 6$\rg] {
\includegraphics[width=55mm]{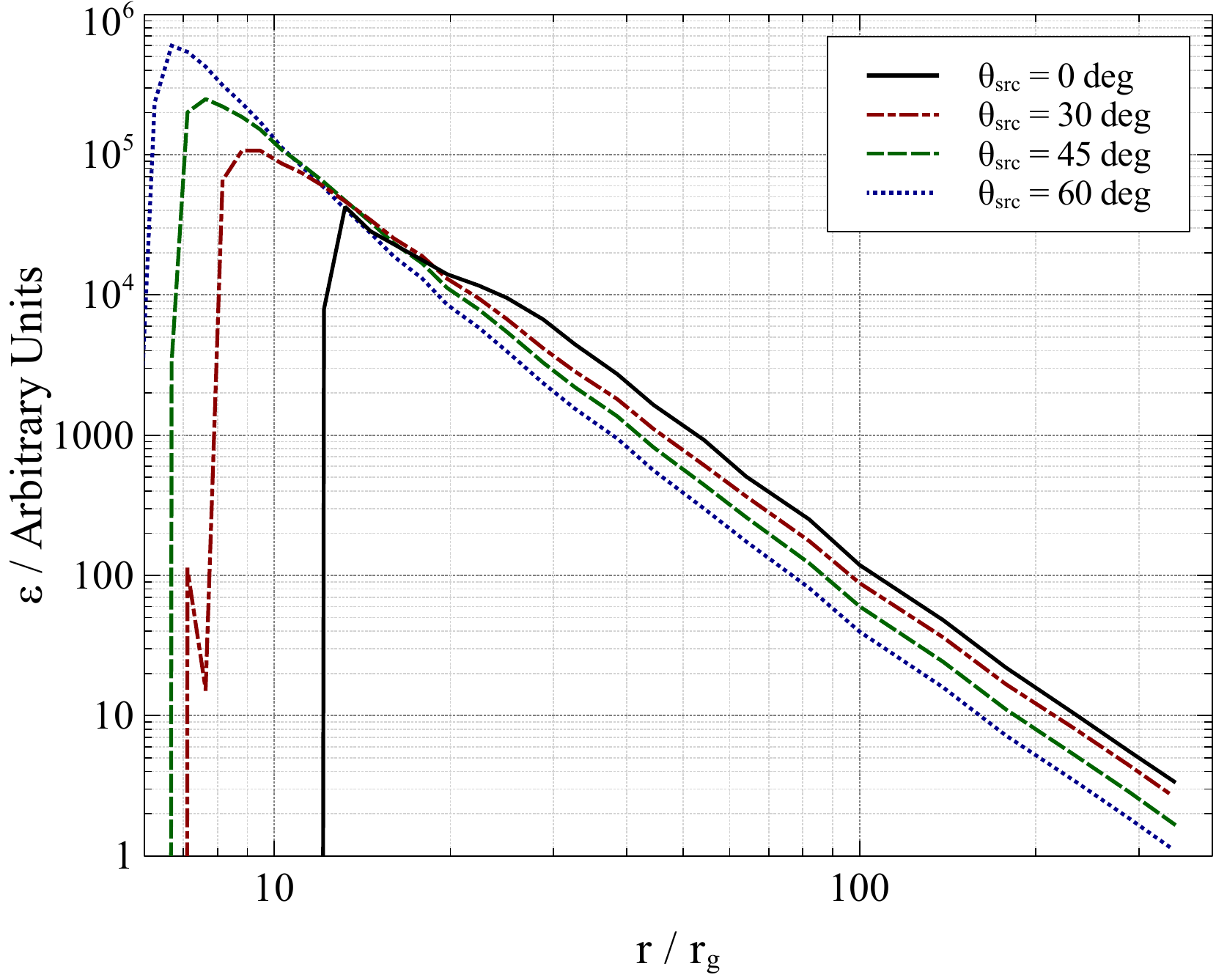}
\label{emis_theta.fig:r6}
}
\caption{Emissivity profiles of accretion discs illuminated by point sources at varying latitude on the neutron star surface for a star of radius \subref{emis_theta.fig:r45} 4.5\rg, \subref{emis_theta.fig:r5} 5\rg\ and \subref{emis_theta.fig:r6} 6\rg\ demonstrating the apparent truncation of the inner edge of the disc when it is illuminated by a point source at high latitude on the neutron star surface.}
\label{emis_theta.fig}
\end{figure*}

We note the apparent truncation of the accretion disc in the case larger neutron stars with a polar ($\theta_\mathrm{src} = 0$) X-ray source. To explore this effect further, emissivity profiles were computed for sources at a range of latitudes on the star surface for three different star radii, shown in Fig.~\ref{emis_theta.fig}. The apparent truncation of the disc occurs because during the ray tracing simulations, rays were not allowed to pass through the surface of the star. For a source at a given latitude, this means that only outward propagating rays were permitted for which $\dot{r}>0$ and that the tangent to each point on the sphere defines the emission boundary. This effectively casts a shadow over the innermost portion of the accretion disc, though only becoming apparent for the highest latitude sources ($\theta_\mathrm{src} < 10$\,deg) for all but the largest star radii. The disc can appear to be truncated to between $7\sim 11$\rg\ for stars of radius 6\rg\ but only to 7\rg\ (with the disc intrinsically extending in to 6\rg) for 4.5\rg\ radius neutron stars.

Fitting the emissivity profiles with a power law function reveals that there is slight variation in the slope, dependent on the latitude of the X-ray source upon the neutron star surface. Considering the star of radius 4\rg with the source located on the pole of the star ($\theta_\mathrm{src} = 0$), the shadowing of parts of the accretion disc by the the star surface causes the apparent emissivity index to increase to $q=3.20$, decreasing towards the classical value of $q=3$ when the source is located at $\theta_\mathrm{src} = 45$\,deg but then steepening again for sources close to the disc plane. When the source is located at $\theta_\mathrm{src} = 80$\,deg, the projection of rays onto the nearby inner regions of the accretion disc increases the emissivity index again to $q=3.22$ for a star of radius 4.5\rg, but as high as $q=3.33$ where the star surface is at radius 6\rg, right up against the inner orbit of the accretion disc.

\subsection{Illumination by the Stellar Surface}
\begin{figure}
\centering
\includegraphics[width=85mm]{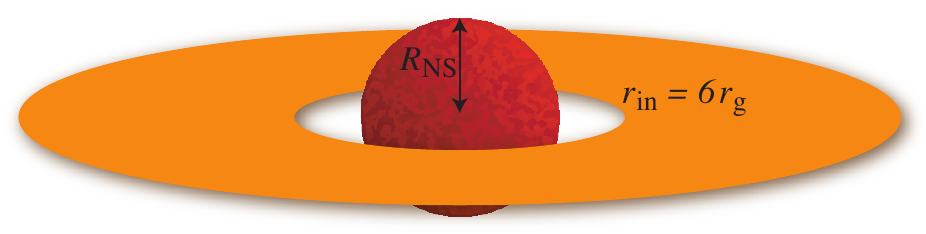}
\caption{The geometry assumed in simulations where the accretion disc is illuminated by the hot neutron star surface.}
\label{hotsurface.fig}
\end{figure}

Illumination by a spherical shell of emission respresenting the surface of the neutron star (Fig.~\ref{hotsurface.fig}) is computed using a Monte Carlo model. A large number of rays are started at random locations upon the shell, travelling in random directions, again to produce isotropic emission from every point on the star surface. No rays are allowed to propagate through the star surface, hence rays are only started travelling outwards ($\alpha \leq \pi/2$) and any ray that propagates into the surface is stopped and not counted towards the emissivity profile.

\begin{figure}
\centering
\includegraphics[width=85mm]{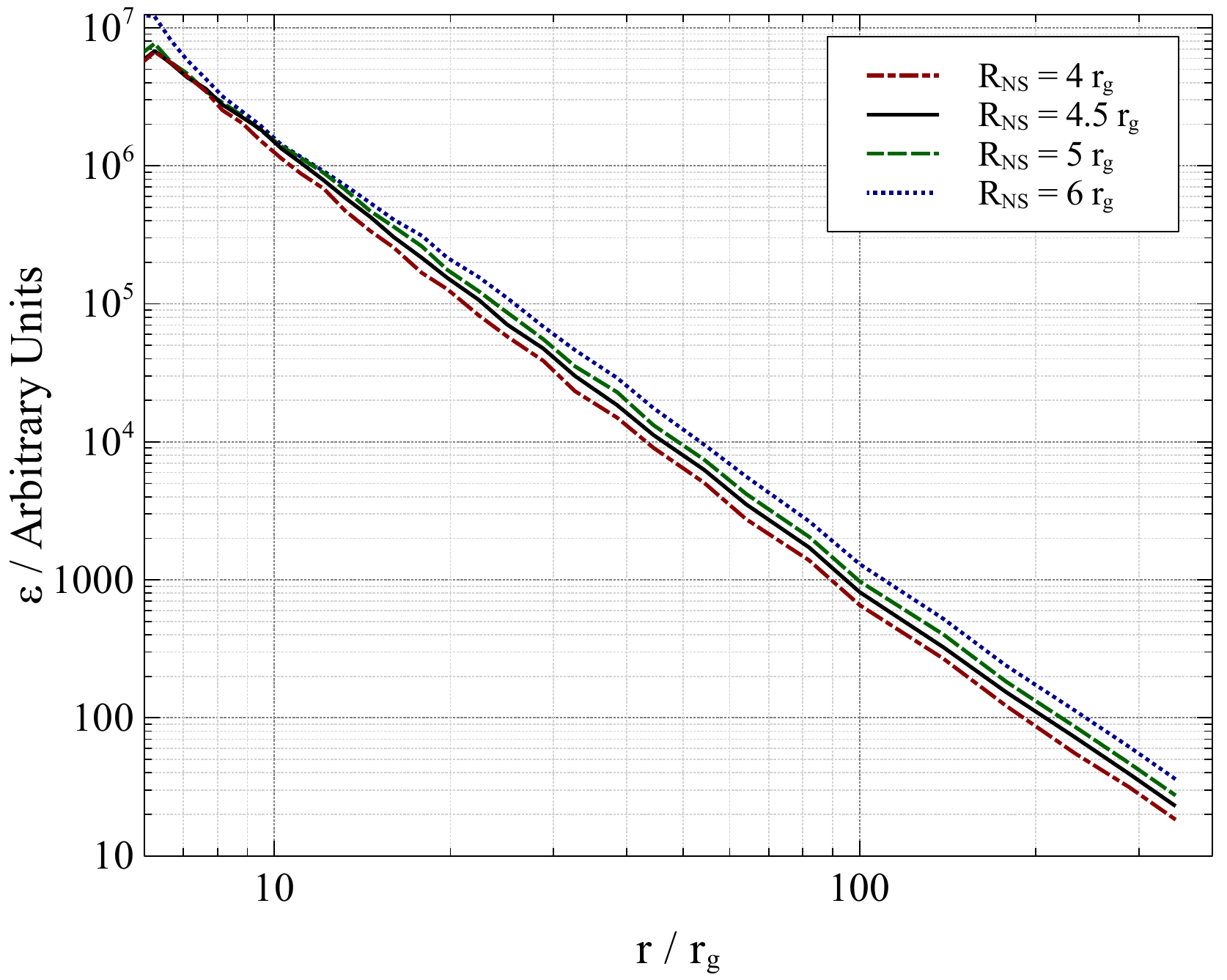}
\caption{Expected emissivity profiles of neutron star accretion discs illuminated by the hot surface of the star. The emissivity profile is shown for neutron stars of varying radius.}
\label{emis_shell.fig}
\end{figure}

Fig.~\ref{emis_shell.fig} shows the calculated emissivity profile of an accretion disc illuminated by a uniform, hot, X-ray emitting surface of a neutron star. Once again, the emissivity profile is consistent with a constant power law index, approximately $q=3$ over the entire disc. For a neutron star of radius 4.5\rg, the best-fitting power law slope to the simulated emissivity profile is $q=3.17$, flattening very slightly to $q=3.12$ when the stellar radius is increased to 6\rg.

The truncation of the disc at 6\rg\ for a slowly rotating neutron star means that the star surface at 4.5\rg\ is sufficiently far from the disc that it essentially appears as a point source to the disc. The constant power law slope is maintained even for larger neutron star radii with only a very subtle steepening over the inner $1\sim 2$\rg\ when the stellar surface is right up against the inner edge of the disc at 6\rg. From the emissivity profile alone it would be difficult to distinguish the case of illumination by a point source and by the entire surface of the star.

\subsection{Illumination by a Disc Corona}
If the accretion disc is strongly magnetised, a corona of energetic particles could be produced over its surface that produce the illuminating X-ray continuum \citep[\textit{e.g.}][]{galeev+79,haardt+91,liu+03,uzdensky+08}. Such a corona extending at low height for tens of gravitational radii over the surface of the disc has been found associated with the accretion discs in a growing number of Seyfert galaxies \citep{understanding_emis_paper}, particularly in their more X-ray luminous phases \citep{1h0707_var_paper,mrk335_corona_paper}.

If such a corona were to provide a significant portion of the disc illumination in accreting neutron star systems, the emissivity profile would follow the same form as it does in black hole systems. The emission directly over the accretion disc causes the emissivity profile to flatten (with power law index $q=0$) in this region then breaking to a constant power law index $q=3$ at the radius coinciding with the edge of the corona. In the case of a neutron star disc, due to the truncation of the accretion disc either at or above the innermost stable circular orbit for spin parameter $a=0$, the inner steepened portion of the emissivity profile would not be seen since this highly-irradiated portion of the inner disc only exists when it extends within a radius of 4\rg\ \citep{1h0707_emis_paper}.

\section{Illumination of the disc in Ser X-1}
Serpens X-1 (Ser~X-1) is one of the best-known neutron star low mass X-ray binary (LMXB), discovered in 1965 \citep{friedman+67}. Being it a bright, persistent X-ray source, it has been well studied by that vast majority of X-ray missions including \textit{NuSTAR} in 2013. \textit{NuSTAR} observations of Ser X-1 reveal, for the first time, the Compton back-scattering `hump' around 20\keV; an important signature of the reflection of X-rays off the accretion disc, in addition to a definitive detection of a relativistically broadened iron K$\alpha$ fluorescence line arising from the disc material orbiting in the strong gravity environment around the neutron star \citep[][henceforth M13]{miller+2013}.

In light of the theoretical models developed above, the illumination of the accretion disc around the neutron star in Ser X-1 was studied from the profile of the relativistically broadened iron K$\alpha$ line detected by \textit{NuSTAR}. Data from the two \textit{NuSTAR} observations (OBSIDs 30001013002 and 30001013004) were reduced following the standard procedure with \textsc{nupipeline} using the most recent calibration database (\textsc{caldb}) at the time of writing and spectra were extracted using \textsc{nuproducts}. Following M13, the spectra from the two focal plane module detectors, FPMA and FPMB were combined across both observations, creating a single summed spectrum.

\begin{figure}
\centering
\includegraphics[width=85mm]{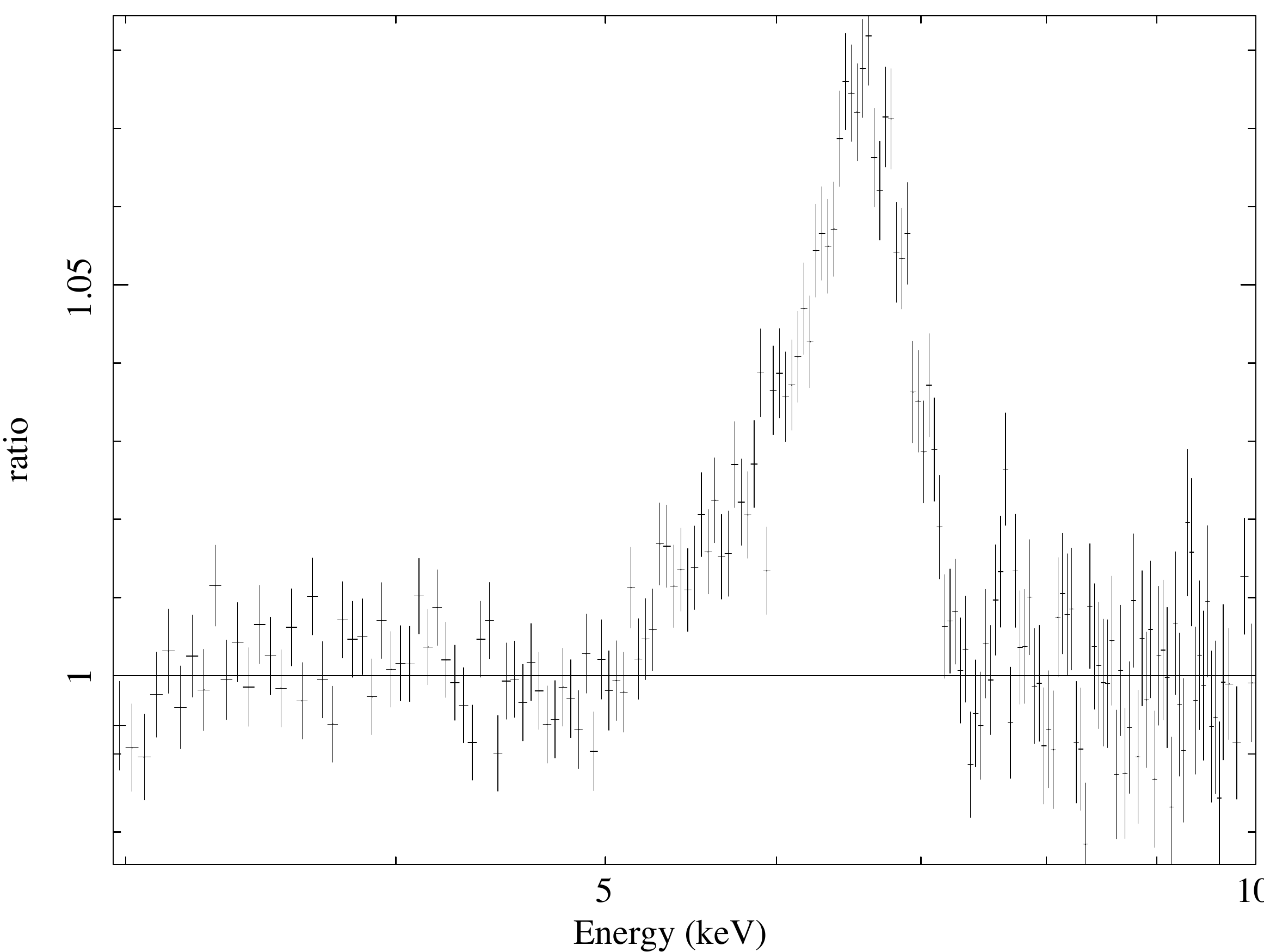}
\caption{The profile of the broadened iron K$\alpha$ from the accretion disc in Ser X-1, shown as the ratio of the summed \textit{NuSTAR} FPMA and FPMB spectra to the best-fitting continuum model in the 3-10\keV\ band that is used for measuring the emissivity profile of the accretion disc.}
\label{serx1_line.fig}
\end{figure}

The \textit{NuSTAR} spectrum was initially fit in \textsc{xspec} \citep{xspec} with a model to describe the continuum underlying the iron K lines in the 3-10\keV\ band. The model was based upon that of M13, consisting of black body emission from the hot surface of the star or the boundary layer in which angular momentum of material from the inner radii of the disc is dissipated in order to accrete onto the more slowly rotating star surface, a multi-temperature black body spectrum representing the thermal emission from the accretion disc and a power law X-ray continuum component. Fig.~\ref{serx1_line.fig} shows the iron K line as residuals to this model. The accretion disc was found to have a best fitting temperature of $(1.80_{-0.02}^{+0.03})$\keV\ with the normalisation of the disc black body component found to be $26_{-1.5}^{+1.3}$. The black body emission from the star has best fitting temperature of $(2.50_{-0.03}^{+0.04})$\keV\ with normalisation $(2.00_{-0.7}^{+0.07})\times 10^{-2}$. The power law X-ray continuum has best fitting photon index $\Gamma = 3.43_{-0.06}^{+0.08}$ and normalisation $1.5\pm0.2$.

M13 provide a complete description of the continuum fitting and show that in order to describe the continuum above 10\keV\ it is necessary to include the reflection from the disc to model the `Compton hump' around 20\keV, though this component is not required here since we only seek a model  to provide the emission underlying the iron K$\alpha$ line which we wish to isolate and determine the precise profile. The reflected continuum component (distinct from the reflected line) does not contribute significantly between 3 and 10\keV.

The line was initially fit using \textsc{laor} model with a single power law emissivity profile in order to determine the best fitting rest frame energy, $E=(6.77^{+0.09}_{-0.09})$\keV, and inclination of the accretion disc to the line of sight, $i=(25.0^{+2.1}_{-1.8})$\,deg. The model consisting of the continuum and the simple line profile gives $\chi^2 / \nu = 1.2$, showing a reasonable though not formally acceptable agreement with the data. Including an absorption edge between 9 and 10\keV, consistent with Fe\,\textsc{xxvi} in the accretion disc significantly improves the fit, yielding $\chi^2 / \nu = 1.1$, showing that the model provides an appropriate description of the continuum.

The emissivity profile of the accretion disc was then measured following the procedure of \citet{1h0707_emis_paper}. The observed broad emission line was fit as the sum of the contributions from successive radii in the disc. The best-fitting normalisation of the contribution to the line from each annulus reveals the illumination profile of the accretion disc by the illuminating X-ray source. Only the line provides statistical constraints on the emissivity profile since this is a narrow feature in the rest frame of the disc material, meaning the range of redshifts that identify the radii of line emission can be determined. Broad spectral features such as the Compton hump in the reflection spectrum or the soft excess where multiple broad emission lines sit atop one another provide no real constraint on the emissivity profile.

Using this rest frame line energy and inclination, the measured emissivity profile was found to be somewhat unrealistic, with emission clustered between 8 and 40\rg\ and the emission from neighbouring points on the disc fluctuating up and down. Measuring the emissivity profile can be a powerful discriminator of the assumed rest frame reflection spectrum. If the incorrect rest frame spectrum is assumed, the model will correct for the shape of the observed spectrum by adjusting the emissivity from each part of the disc to produce the required emission at each energy over the range of redshifts that are observed in the line. This results in the measurement of an emissivity profile that does not make sense in the context of the illumination of the disc by an X-ray source with sharp peaks and dips and/or an abrupt cut-off before the illumination reaches the outer disc.

\begin{figure}
\centering
\includegraphics[width=85mm]{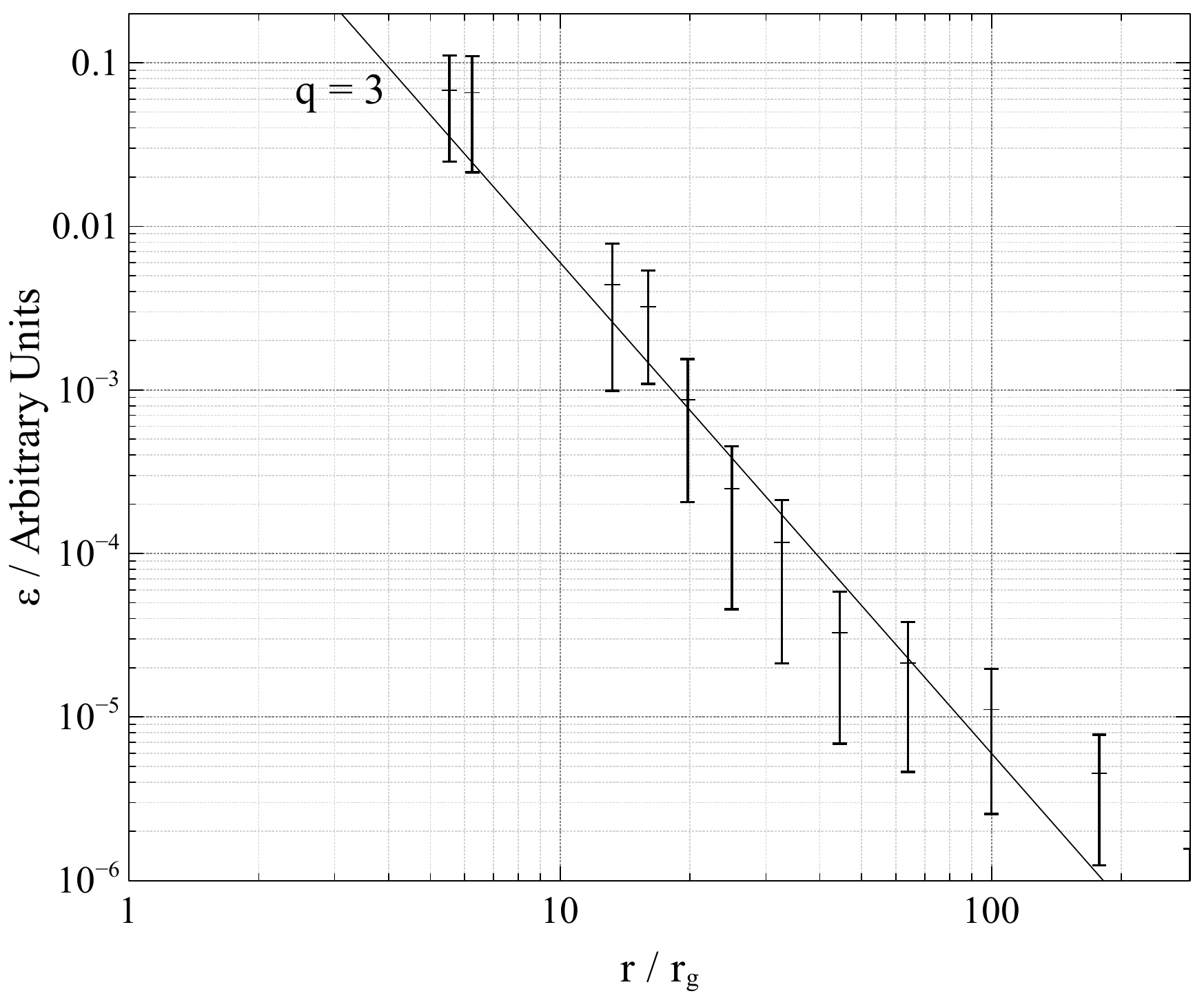}
\caption{The emissivity profile of the accretion disc around the neutorn star in the LMXB Ser X-1. The emissivity profile is measured following the procedure of \citet{1h0707_emis_paper}, fitting the contribution to the relativistically broadened Fe\,\textsc{xxv} K$\alpha$ line (rest frame energy 6.67\keV) from successive radii in the disc.}
\label{serx1_emis.fig}
\end{figure}

Instead of allowing the rest frame energy of the emission line to vary as a free parameter, the model was fit using a single line (M13 find no evidence for significant contributions to the spectrum by anything more than a single, broadened emission line) at the energy of the K$\alpha$ line of neutral iron (6.4\keV), Fe\,\textsc{xxv} (6.67\keV) and Fe\,\textsc{xxvi} (6.97\keV). Of these three options, the spectrum is best fit by the Fe\,\textsc{xxv} K$\alpha$ line at $E = 6.67$\keV\ (with best-fitting inclination $i=(28.6^{+0.6}_{-0.6}$\,deg) which yields the same goodness-of-fit as when the line energy is allowed to vary freely (indeed, $6.67$\keV\ lies marginally below the 90 per cent confidence limit of the free line energy). The resulting emissivity profile is shown in Fig.~\ref{serx1_emis.fig}.

Line emission is seen from the accretion disc extending right down to the innermost stable circular orbit at 6\rg, although disc emission from between 7 and 12\rg\ on the disc is not well constrained. This is typically seen when fitting the emissivity profile to the profile of relativistically broadened emission lines, as discussed by \citet{1h0707_emis_paper}. The peak of the line is fit by the outer radii of the disc and the redshifted wing is given by the inner radii, however the middle radii of the disc (in this case $7-12$\rg) do not uniquely contribute to any energy, so they are poorly constrained by the data. It should also be noted that while, in principle, the increased count rate from X-ray binaries allows for more precise determination of the profile of broadened emission lines, in practice this is hindered by the relatively weak line compared to the bright thermal X-ray emission from the neutron star surface and the accretion disc and the fact that the fluorescent yield is less from the more highly ionised Fe\,\textsc{xxv} ions indicated by the 6.67\keV\ centroid energy of the line (compared to 6.4\keV\ for neutral iron).

The measured emissivity profile shows an almost constant power law slope as a function of radius. For reference, the canonical $r^{-3}$ power law is over-plotted. While being consistent with a single slope $q=3$, the best-fitting power law slope to the measured data points is $q=3.5_{-0.4}^{+0.3}$, consistent with that of the single emission line fit over the entire disc by M13. The emissivity index being measured around 3.5 along with there being no significant residuals between the observed spectrum and the model corroborate the description of the emission line with an axisymmetric emissivity profile.

From these relatively simple measurements of the iron K$\alpha$ line profile in Ser X-1 and the inferred illumination pattern of the accretion disc by the primary X-ray source, it is possible to begin to place some constraints upon the nature and geometry of the X-ray continuum source in this neutron star LMXB. Firstly, the detection of an approximately constant power law slope in the emissivity profile over the inner radii of the disc with no evidence for flattening of the profile suggests that the illumination of the disc is at least dominated by a source associated with and concentrated around the neutron star. There is no evidence for a significant contribution to the non-thermal X-ray emission (that excites the fluorescence line) arising from a corona above the accretion disc.

The constraints on the inner radius of the accretion disc from the broad emission line profile with no drop in the emission line emissivity being seen within 10\rg\ rule out the X-ray source being entirely contained within 30\,deg of the pole of the neutron star, with no apparent truncation of the accretion disc outside the innermost stable circular orbit. A hotspot on the surface of the star would likely be created by the funnelling of the accreting material along magnetic field lines onto the pole of the dipole field. Such a strong magnetic field is expected to disrupt the inner regions of the accretion disc, truncating the disc within the Alfv\'enic radius, where the magnetic energy density equals the rest mass energy \citep{davidson+73,illarionov+75}. Since no such truncation of the accretion disc is seen, it follows that the accreting material is not directed by the magnetic field onto high-latitude hot spots on the neutron star surface.

Having already ruled out high-latitude X-ray sources, the steepness of the best-fitting power law to the emissivity profile suggests that in Ser X-1, the X-ray emitting region is located at high $\theta_\mathrm{src}$, close to the disc plane. Ray tracing simulations have shown that power law indices up to $3.2\sim 3.3$ can be produced where $\theta_\mathrm{src}$ is above 60\,deg. It therefore seems likely that the X-ray continuum emission is arising from the boundary layer. It is, however, not possible to rule out discrete hotspots within this region that rotate with the neutron star surface since the line profile only suggests axisymmetric illumination of the disc once averaged over the observation.

The ray tracing simulations assume an idealised geometry in which the X-ray continuum is emitted isotropically from the surface of the source region and that it illuminates a planar accretion disc. Where X-ray emission arises from an extended boundary layer between the disc and the neutron star, there are likely additional geometric effects including severe limb-darkening of the emission. The parts of the boundary layer closer to the disc plane will predominantly illuminate the innermost regions of the disc. On the other hand, illumination reaching the outer parts of the disc will originate from higher latitude regions of the boundary layer. This emission will not travel radially out of the boundary layer, hence will graze the outer surface of the boundary layer. This limb-darkening will reduce the emission reaching the outer disc relative to that hitting the inner disc, which could explain some addition steepening in the observed emissivity profile above what is seen in idealised simulations.

There is some evidence for the flattening of the emissivity profile in Ser X-1 at large radius. Fitting a broken power law, rather than a single power law slope, yields a better fit to the measured emissivity profile. Comparing to the original fit of a single power law, an $F$ statistic of 9.2 is found ($p=0.01$), indicating a significantly better fit to the data. At a break radius of $(44_{-16}^{+24})$\rg, the emissivity profile was found to flatten to a power law index of $1.9_{-0.5}^{+0.2}$. This apparent flattening of the emissivity profile could be a `wiggling' of the profile that is caused by slight inadequacies in the modelling of the rest-frame reflection spectrum of the accretion disc. Although M13 find that only a single broad emission line is required in the iron K band, there could be slight contributions to this line from other ionisation states of iron that produce subtle changes in the line profile that are fit as structure in the emissivity profile. There is not sufficient signal to noise in the detection of the line above the continuum in these observations of Ser X-1 to rigorously test the underlying emission line structure.

\section{Conclusions}

Ray tracing simulations from either point-like hotspots, bands of emission or from the spherical surface of a neutron star show that the illumination of the accretion disc by X-ray continua arising from these sources is well described by a single power law emissivity profile with the index slightly steeper than the classical case $r^{-3}$.

Unless the neutron star radius is larger than the innermost stable orbit ($\gtrsim 6$\rg), or the disc is viewed at inclination greater than 60\,deg, the emission line seen from an axisymmetrically illuminated accretion disc is well described by the canonical suite of relativistically broadened emission line models commonly applied to black hole accretion flows. At higher inclination or greater neutron star radius, shadowing of the back side of the disc can lead to a reduction in line flux from the inner disc by around 10 per cent, although in the most extreme cases up to 35 per cent. The inner radius of the disc, the emissivity profile and the inclination can still be well determined, however, using the black hole line models.

If the emitting region is at low latitude, close to the accretion disc, geometric effects cause steepening of the emissivity profile power law index as high as $\sim 3.3$ while a high latitude source, close to the pole of the neutron star, can lead to the apparent truncation of the accretion disc as the inner regions of the disc are shadowed from X-ray continuum emission by the opaque neutron star surface.

Where the accretion disc is illuminated by a hotspot on a neutron star that rotates on longer timescales than the integration time of the observation, non-axisymmetric illumination of the accretion disc due to the shadowing of one side of the accretion disc by the star results in skewed emission line profiles. In these cases, the observed line is not well described by the canonical models and can lead to either the measurement of a steeply falling emissivity profile, falling off as steep as $r^{-7}$, or underestimating the inclination of the accretion disc.

Measuring the emissivity profile of the accretion disc in the LMXB Serpens X-1 reveals a profile consistent with a single power law over the accretion disc. No evidence is seen for truncation of the accretion disc outside the innermost stable orbit in the Schwarzschild spacetime and the emissivity index is measured at $q=3.5_{-0.4}^{+0.3}$. No evidence is seen for flattening of the emissivity profile over the inner accretion disc suggesting that there is no significant contribution to the X-ray continuum emission that illuminates the disc from a corona extending over the surface of the disc. The accretion disc is predominantly illuminated by X-rays emitted from close to the neutron star itself.

The steepness of the observed emissivity profile is not fully explained by simple theoretical models of the X-ray propagation but is suggestive of the X-ray continuum predominantly arising from a narrow belt at low latitude, close to the plane of the accretion disc. The enhanced steepening of the profile could be explained by limb darkening of the emission from a tenuous region, hence it is plausible that the illuminating X-ray continuum arises from the boundary layer in which the rapidly orbiting material on the inner edge of the disc must decelerate to accrete onto the more slowly rotating neutron star.

\section*{Acknowledgements}
DRW is supported by NASA through Einstein Postdoctoral Fellowship grant number PF6-170160, awarded by the \textit{Chandra} X-ray Center, operated by the Smithsonian Astrophysical Observatory for NASA under contract NAS8-03060. Thanks must go to Jon Miller for the insightful discussions that led to this work and to the anonymous referee whose feedback improved the structure of this manuscript. This work used the XStream computational resource, supported by the National Science Foundation Major Research Instrumentation program (ACI-1429830).

\bibliographystyle{mnras}
\bibliography{agn}

\label{lastpage}

\end{document}